\documentclass[final,5p,times,twocolumn]{elsarticleMod}
\usepackage{amsmath,amsfonts,amssymb,array}
\usepackage{bm,bbm,bbm,bbold}
\usepackage{enumerate}
\usepackage{dsfont}
\usepackage{float}
\usepackage{graphicx}
\usepackage{longtable}
\usepackage{multirow}
\usepackage{natbib}
\usepackage{slashed,subfigure}
\usepackage{xcolor,xspace}
\usepackage[colorlinks = true,linkcolor = blue]{hyperref}
\hypersetup{bookmarks=true,unicode=true,pdftoolbar=true,pdfmenubar=true,
  pdffitwindow=false,pdfstartview={FitH},
  pdftitle={Anatomy of inclusive ttW production - [arXiv:2009.00032]}, pdfauthor={von Buddenbrock, Ruiz, Mellado},
  pdfsubject={Subject},pdfcreator={Creator},pdfproducer={Producer},
  pdfkeywords={Top Quarks}{Jets}{Large Hadron Collider}
}
\def\fb{{\rm ~fb}}

\newcommand{\invfb}{{\rm ~fb^{-1}}}

\def\GeV{{\rm ~GeV}}
\def\TeV{{\rm ~TeV}}
\newcommand{\ttW}{\ensuremath{t\bar{t}W}\xspace}
\newcommand{\ttWj}{\ensuremath{t\bar{t}W}j\xspace}
\newcommand{\ttWjj}{\ensuremath{t\bar{t}W}jj\xspace}
\newcommand{\ttWpm}{\ensuremath{t\bar{t}W^\pm}\xspace}
\newcommand{\ttWpmj}{\ensuremath{t\bar{t}W^\pm j}\xspace}
\newcommand{\ttWpmjj}{\ensuremath{t\bar{t}W^\pm jj}\xspace}

\newcommand{\ttZ}{\ensuremath{t\bar{t}Z}\xspace}

\newcommand{\twolep}{\texttt{2lSS}\xspace}
\newcommand{\threelep}{\texttt{3l}\xspace}
\newcommand{\mgamc}{\texttt{mg5amc}\xspace}

\journal{CP3-20-41, MCnet-20-20, VBSCAN-PUB-08-20, ICPP-031} % see changes at L450 in elsarticleMod.cls
\begin{document}
\begin{frontmatter}
%%%%%%%%%%%%%%%%%%%%%%%%%%%%%%%%%%%%%%%
%%%%%%%%%%%%%%%%%%%%%%%%%%%%%%%%%%%%%%%
\title{Anatomy of inclusive $t\overline{t}W$ production at hadron colliders}

\author[Wits]{Stefan von Buddenbrock}  
\ead{stef.von.b@cern.ch}

\author[CP3]{Richard Ruiz}  
\ead{richard.ruiz@uclouvain.be}

\author[Wits,iThemba]{Bruce Mellado}  
\ead{Bruce.Mellado.Garcia@cern.ch}

\address[Wits]{School of Physics and Institute for Collider Particle Physics, University of the Witwatersrand, Wits, Johannesburg 2050, South Africa}
\address[CP3]{Centre for Cosmology, Particle Physics and Phenomenology {\rm (CP3)}, 
Universit\'e catholique de Louvain, Chemin du Cyclotron, Louvain-la-Neuve, B-1348, Belgium}
\address[iThemba]{iThemba LABS, National Research Foundation, PO Box 722, Somerset West 7129, South Africa}

%This article is registered under preprint number: arXiv:2009.00032
%%%%%%%%%%%%%%%%%%%%%%%%%%%%%%%%%%%%%%%
%%%%%%%%%%%%%%%%%%%%%%%%%%%%%%%%%%%%%%%
\begin{abstract}
In LHC searches for new and rare phenomena the top-associated channel $pp \to t\overline{t}W^\pm +X$ is a challenging background that multilepton analyses must overcome. Motivated by sustained measurements of enhanced rates of same-sign and multi-lepton final states,  we reexamine the importance of higher jet multiplicities in $pp \to t\overline{t}W^\pm +X$ that enter at $\mathcal{O}(\alpha_s^3\alpha)$ and $\mathcal{O}(\alpha_s^4\alpha)$,  i.e.,  that contribute at NLO and NNLO in QCD in inclusive  $t\overline{t}W^\pm$ production. Using fixed-order computations, we estimate that a mixture of real and virtual corrections at $\mathcal{O}(\alpha_s^4\alpha)$ in well-defined regions of phase space  can arguably increase the total  $t\overline{t}W^\pm$ rate at NLO by at least $10\%-14\%$. However, by  using non-unitary  NLO multi-jet matching, we estimate  that these same corrections are at most  $10\%-12\%$, and at the same time exhibit the enhanced jet multiplicities that are slightly favored by data. This seeming incongruity suggests a need  for the full NNLO  result. We comment  on implications for the $t\overline{t}Z$ process.
\end{abstract}

\begin{keyword}
{\small
Top  Quarks  \sep Jet Matching \sep  Large Hadron Collider %\sep arXiv:2009.00032
}
\end{keyword}

\end{frontmatter}

%%%%%%%%%%%%%%%%%%%%%%%%%%%%%%%%%%%%%%%%%%%%%%%%%%%%%%%%%%%%%%%%%%%%%%%%%%%%%%
\section{Introduction}\label{sec:Intro}
%%%%%%%%%%%%%%%%%%%%%%%%%%%%%%%%%%%%%%%%%%%%%%%%%%%%%%%%%%%%%%%%%%%%%%%%%%%%%%

\begin{figure}
    \centering
    \includegraphics[width=.9\columnwidth]{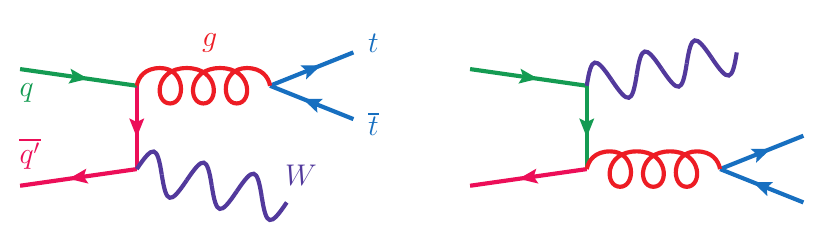}
    \caption{Lowest order, Born-level diagram for  the $q\overline{q}\to t\overline{t}W^\pm$ process.}
    \label{fig:ttWAnatomy_diagram_ttW_Born}
\end{figure}

The discovery \cite{Aad:2015eua,Khachatryan:2015sha} of the  $pp\to \ttW$ process, and likewise $pp\to\ttZ$, is an important milestone of the Large Hadron Collider's (LHC's)  Standard Model (SM), Higgs, and  New Physics programs. In its own right  \ttW, which at lowest order  proceeds at $\mathcal{O}(\alpha_s^2 \alpha)$ through the diagrams in figure~\ref{fig:ttWAnatomy_diagram_ttW_Born}, is a multi-scale process with large quantum chromodynamic (QCD)  and electroweak (EW) corrections. Hence, it  is a laboratory for stress-testing the SM paradigm. At the same time, the $\ttWpm \to W^+W^-W^\pm b\overline{b}$ decay mode can give rise to the same-sign dilepton $\ell^\pm_i \ell^\pm_j$ and trilepton $\ell_i \ell_j \ell_k$ signal categories, encumbering~\cite{delAguila:2008cj,Maltoni:2015ena,deFlorian:2016spz,Pascoli:2018heg}  searches for lepton number and lepton flavor violation as well as measurements of the Higgs's couplings.

Motivated by this, major efforts have been undertaken since the top's discovery to reliably describe the $\ttW/Z$  processes. This includes QCD corrections to production and decay modes up to next-to-leading order (NLO) with parton shower (PS) matching \cite{Maltoni:2015ena,FebresCordero:2006nvf,Badger:2010mg,Lazopoulos:2008de,Kardos:2011na,Campbell:2012dh,Garzelli:2011is,Hoeche:2012yf,Garzelli:2012bn,Alwall:2014hca,Maltoni:2014zpa,Tsinikos:2017,Denner:2020hgg,Bevilacqua:2020pzy}; soft gluon resummation up to next-to-next-to-leading logarithm (NNLL) in perturbative QCD \cite{Kulesza:2017hoc,Kulesza:2018tqz,Kulesza:2020nfh} and effective field theory  \cite{Li:2014ula,Broggio:2016zgg,Broggio:2017kzi,Broggio:2019ewu}; EW corrections up to NLO  \cite{Broggio:2019ewu,Frixione:2014qaa,Frixione:2015zaa,Dror:2015nkp,Frederix:2017wme,Frederix:2020jzp},
as well as their systematic merger with QCD corrections~\cite{Broggio:2019ewu,Frixione:2015zaa,Frederix:2020jzp,Frederix:2018nkq}.

The findings are telling: known QCD corrections increase total LHC rates by {$20\%-60\%$}, depending on  theoretical inputs, and reflect the {$15\%-85\%$} scale ambiguity at leading order (LO). However, even at this level, typical scale choices leave an {$10\%-30\%$} uncertainty, suggesting additional corrections are needed to ensure theoretical control. While EW corrections increase rates by a net  {$5\%$}, uncertainties essentially stay the same. 

In comparison to data, a consistent picture has also emerged: Whereas first observations of the $\ttZ$ process by the ATLAS and CMS collaborations at $\sqrt{s}=8\TeV$ were within SM expectations at NLO in QCD, both collaborations measured a $\ttW$ rate exceeding  predictions at the 68\% $(1\sigma)$ confidence level (CL) \cite{Aad:2015eua,Khachatryan:2015sha}. Measurements at $\sqrt{s}=13\TeV$ with up to $\mathcal{L}\approx36\invfb$ support a  $\ttW$ rate that is {$15\%-50\%$} larger than predictions at NLO in QCD at about the same CL tension \cite{Sirunyan:2017uzs,ATLAS:2018ekx,Aaboud:2019njj}; a modest $\ttZ$ rate increase of  {$15\%$} is also preferred~\cite{Sirunyan:2017uzs}. Improved measurements with $\mathcal{L}\approx80\invfb$ \cite{ATLAS:2019nvo} and $\mathcal{L}\approx140\invfb$~\cite{ATLAS:2020hrf,CMS:2020iwy} affirm a $\ttW$ rate that, depending on the signal category, is {$25\%-70\%$} larger than predictions at NLO in QCD and EW,  and corresponds to a {$1.4\sigma-2.4\sigma$} discrepancy. Given the sustained nature of these and other multilepton excesses, which include a diverse number of final states that are dominated by several SM processes, it is reasonable to contemplate seriously the possible role of new physics~\cite{vonBuddenbrock:2016rmr,vonBuddenbrock:2017gvy,vonBuddenbrock:2019ajh,Hernandez:2019geu}. That said, it is also necessary to investigate each anomaly separately  to understand the possible importance of missing higher order corrections. 

To support such investigations and to exhaust possible SM explanations we have reexamined the role of the $t\overline{t}W^\pm j$ and $t\overline{t}W^\pm j  j$ sub-processes at $\mathcal{O}(\alpha_s^3 \alpha)$ and $\mathcal{O}(\alpha_s^4 \alpha)$ in inclusive $\ttW$ production. While complementary works have studied the phenomenology of these channels~\cite{Maltoni:2015ena,Alwall:2014hca,Tsinikos:2017}, the impact on the inclusive cross section were not among their intents.

As a first step we use fixed-order computations and find that a subset of well-defined real and virtual contributions at $\mathcal{O}(\alpha_s^4 \alpha)$, i.e., finite elements to inclusive $\ttW$ production at next-to-next-to-leading order (NNLO) in QCD, are positive and reach at least {$10\%-14\%$} of the $t\overline{t}W$ rate at NLO. Interestingly, we find that these same contributions only increase the $\ttW$ rate  by at most {$10\%-12\%$} when using then the non-unitary, NLO multi-jet matching procedure FxFx \cite{Frederix:2012ps}. Despite this seeming discrepancy, we report that after imposing  selection cuts and signal categorizations employed~\cite{ATLAS:2019nvo} by LHC experiments, the FxFx results exhibit enhanced light and heavy jet multiplicities that are slightly favored by data. To resolve this enigma, we argue a need for the full NNLO in QCD description of inclusive $\ttW$ production.

The report of our investigation continues as follows: After summarizing our computational setup in section~\ref{sec:mc}, we build up the anatomy of inclusive $\ttW$ production at hadron colliders in section~\ref{sec:inclusive}.
There we estimate higher order corrections to $\ttW$ production  and discuss theoretical uncertainties. In section~\ref{sec:differential}, we show how rate increases propagate to differential observables and survive analysis cuts. In section~\ref{sec:outlook}, we present an outlook for the $\ttZ$ process.
We conclude in section~\ref{sec:conclusions}.

%%%%%%%%%%%%%%%%%%%%%%%%%%%%%%%%%%%%%%%%%%%%%%%%%%%%%%%%%%%%%%%%%%%%%%%%%%%%%%
\section{Computational Setup}\label{sec:mc}
%%%%%%%%%%%%%%%%%%%%%%%%%%%%%%%%%%%%%%%%%%%%%%%%%%%%%%%%%%%%%%%%%%%%%%%%%%%%%%

To conduct our study we employ a state-of-the-art simulation tool chain based on Monte Carlo methods. For matrix element evaluation and parton-level event generation, we use   \texttt{MadGraph5\_aMC@NLO} (v2.6.7) \cite{Alwall:2014hca} (\mgamc). Its conglomeration of  packages \cite{Frixione:2002ik,Frederix:2009yq,Hirschi:2011pa,Artoisenet:2012st,Alwall:2014bza,Hirschi:2015iia} enables us to simulate high-$p_T$ hadron collisions in the SM up to NLO in QCD with PS-matching  within  the MC@NLO formalism~\cite{Frixione:2002ik}. We model decays of heavy resonances using the spin-correlated narrow width approximation   \cite{Artoisenet:2012st,Alwall:2014bza}. Parton-level events are passed through \texttt{Pythia8} (v244) \cite{Sjostrand:2014zea} for QCD and QED parton showering, hadronization, and modeling of the underlying event.  We use the FxFx prescription~\cite{Frederix:2012ps}  as implemented in \mgamc. Parton-level sequential clustering proceeds according to   $k_T$-class  algorithms~\cite{Catani:1993hr,Ellis:1993tq,Cacciari:2008gp} as implemented in \texttt{FastJet} \cite{Cacciari:2005hq,Cacciari:2011ma}. To compare against ATLAS $\ttW$ results at $\sqrt{s}=13\TeV$~\cite{ATLAS:2019nvo},  events are processed with \texttt{DELPHES3} (v3.4.2) \cite{deFavereau:2013fsa} to model detector resolution. We assume most of the default settings for the ATLAS detector card. However, to better mimic ATLAS's  analysis we employ updated lepton and $b$-tagging efficiencies \cite{Aad:2019tso,Aad:2016jkr,Aad:2019aic}.

Throughout this analysis, we work in the $n_f=4$ active quark flavor scheme with SM inputs set by the \mgamc module \texttt{loop\_sm}. We do so for a more realistic description of massive ${B}$ hadrons decays, and particularly charged lepton multiplicities.
We have checked that this results in NLO and FxFx cross section normalizations that are about $6\%$ larger than in the  $n_f=5$ scheme.
We tune the top and Higgs masses to
\begin{equation}
m_t(m_t) = 172.9\GeV \quad\text{and}\quad m_H = 125.1\GeV.
\end{equation}
For all computations we use the NNPDF 3.1 NNLO parton density functions (PDFs) (\texttt{lhaid=303600})~\cite{Ball:2017nwa}  as evaluated by \texttt{LHAPDF} (v6.2.3)~\cite{Buckley:2014ana}. PDF uncertainties are obtained via the replica method \cite{Buckley:2014ana,Ball:2017nwa}. We take our central $(\zeta=1)$ collinear factorization $(\mu_f)$ and QCD renormalization $(\mu_r)$ scales to be half the sum over final-state transverse energies,
\begin{equation}
\mu_f, \mu_r = \zeta \times \frac{\tilde{H}_T}{2}, \quad \tilde{H}_T \equiv \sum_{k =  t,\overline{t},W^\pm,j} \sqrt{m_k^2 + p_{T,k}^2},
\label{eq:scale}
\end{equation}
where $m_f$ and $p_{T,k}$ are the mass and transverse momentum of final-state particle $k$.
The shower scale $\mu_s$ is kept at its default value \cite{Alwall:2014hca}. 
For FxFx computations, scales are set according to Refs. \cite{Frederix:2012ps,Alwall:2014bza}.
Uncertainties associated with $\mu_f, \mu_r,$ and $\mu_s$ are estimated by rescaling them individually by $\zeta\in[0.5,1,2.0]$.

%%%%%%%%%%%%%%%%%%%%%%%%%%%%%%%%%%%%%%%%%%%%%%%%%%%%%%%%%%%%%%%%%%%%%%%%%%%%%%
\section{Anatomy of inclusive \ttW production at the LHC}\label{sec:inclusive}
%%%%%%%%%%%%%%%%%%%%%%%%%%%%%%%%%%%%%%%%%%%%%%%%%%%%%%%%%%%%%%%%%%%%%%%%%%%%%%

It may be that sustained measurements \cite{Sirunyan:2017uzs,ATLAS:2018ekx,Aaboud:2019njj,CMS:2020iwy,ATLAS:2019nvo,ATLAS:2020hrf} of a $\ttW$ cross section that is larger than expectations at NLO in QCD and EW is due to new physics. As such, we find it compelling to exhaust  SM explanations for these observations. In this context, we review in section \ref{sec:inclusive_state} the modeling and uncertainties  of $\ttW$ production at various orders in perturbation theory. Motivated by our findings,  we turn our  focus in section \ref{sec:inclusive_ttwjx}  to the $\ttWj$ and $\ttWjj$ sub-processes, and their roles in the inclusive channel. We then present in section \ref{sec:inclusive_fxfx} our estimations for the $\ttW$ production rate at the level of NLO multi-jet matching. Differential results are presented in section \ref{sec:differential}.

%%%%%%%%%%%%%%%%%%%%%%%%%%%%%%%%%%%%%%%%%%%%%%%%%
\subsection{State-of-the-art modeling for inclusive production}\label{sec:inclusive_state}

Categorically, the anatomy of inclusive $\ttW$ production consists of several pieces and nuances. To start: at lowest  order, i.e., at $\mathcal{O}(\alpha_s^2 \alpha)$, cross sections at the $\sqrt{s}=13\TeV$ LHC span $\sigma_{\ttW}^{\rm LO}\sim 375\fb-525\fb$, depending on choices for PDF and $\mu_f,~\mu_r$. The normalization of $\alpha_s(\mu_r)$ heavily influences the outcome. For static scale choices of  $\mu_f,\mu_r \sim \mathcal{O}(m_t)$,   3- or 9-point  scale variation reveals an ambiguity of {$25\%-35\%$} \cite{Campbell:2012dh}. For typical dynamic choices, such as equation~\ref{eq:scale}, one finds comparable uncertainties of {$20\%-30\%$} but rates that are about {$25\%$} smaller  \cite{Alwall:2014hca,Maltoni:2014zpa,Maltoni:2015ena}.
The same holds for static choices of $\mathcal{O}(2m_t+M_W)$, indicating that the threshold and kinematic scales are similar.

At NLO in QCD, contributions at $\mathcal{O}(\alpha_s^3 \alpha)$ improve the picture dramatically. Due to the opening of $(qg)$-scattering, cross sections jump by {$20\%-50\%$}, again depending on inputs, to {$\sigma^{\rm NLO}_{\ttW}\sim485\fb-645\fb$} \cite{Maltoni:2015ena,Campbell:2012dh,Alwall:2014hca,Maltoni:2014zpa,Kulesza:2018tqz}. Excluding scale choices that favor values far below $ \mathcal{O}(2m_t+M_W)$, rates more moderately span {$\sigma^{\rm NLO}_{\ttW}\sim485\fb-595\fb$}. However, this range does not reflect scale variation, which spans only {$10\%-15\%$}, and suggests that higher order corrections are needed to ensure theoretical control.
This is partly due to the $(qg)$-scattering channel, which at this level only is described at LO.

Beyond leading contributions at $\mathcal{O}(\alpha_s^2 \alpha)$ and $\mathcal{O}(\alpha_s^3 \alpha)$, it is now known~\cite{Frixione:2014qaa,Frixione:2015zaa,Dror:2015nkp,Frederix:2017wme} that  ``supposedly'' sub-leading EW contributions at the Born level, i.e., at $\mathcal{O}(\alpha^3)$, and at NLO, i.e., at $\mathcal{O}(\alpha_s^2\alpha^2)$, $\mathcal{O}(\alpha_s\alpha^3)$, and  $\mathcal{O}(\alpha^4)$, are  not negligible  in comparison to the above  uncertainty budget. Cancellations among virtual EW diagrams, interference between mixed EW-QCD and pure EW diagrams, real radiation, and the opening of $t W\to t  W$  scattering, culminate to a positive contribution to  $\ttW$ production that  is about {$6\%$} of the rate at NLO in QCD \cite{Frederix:2017wme}:
\begin{equation}
K_{\rm NLO-EW} = \sigma^{\rm NLO-EW+NLO-QCD}/ \sigma^{\rm NLO-QCD} = 1.06.
\label{eq:kFactEW}
\end{equation}
Despite these improvements, dynamic scale variation at this order remains about the same as at NLO in QCD.

First attempts to extract two-loop predictions through resuming soft gluon radiation in the $q\overline{q}\to \ttW$ channel up to NNLL yield positive corrections \cite{Kulesza:2018tqz,Kulesza:2020nfh,Li:2014ula,Broggio:2016zgg,Broggio:2017kzi,Broggio:2019ewu}. Depending on scale inputs, these range {$1\%-7\%$} and reduce slightly both the scale uncertainty and range of predictions~\cite{Kulesza:2018tqz}.

In one\footnote{Other measurements at $\sqrt{s}=13\TeV$~\cite{Sirunyan:2017uzs,ATLAS:2018ekx,Aaboud:2019njj,CMS:2020iwy,ATLAS:2019nvo,ATLAS:2020hrf} show similar disagreements but provide fewer Monte Carlo modeling details.} detailed comparison to data~\cite{ATLAS:2019nvo}, measurements of the $\ttW$ cross section  by ATLAS  with $\mathcal{L}\approx80\invfb$ at $\sqrt{s}=13\TeV$ find category-based signal strengths that are {$25\%-70\%$} larger than SM  expectations. Relative to the SM prediction of {$\sigma^{\rm ATLAS-TH.}_{\ttW}=727\fb^{+13\%}_{-13\%}$}, the measurements indicate  a best-fit rate and signal strength of~\cite{ATLAS:2019nvo},
\begin{equation}
\hat{\sigma}_{\ttW}^{\rm ATLAS-EX.} = 1010\fb~^{+12\%}_{-12\%},
\quad
\hat{\lambda}_{\ttW}^{\rm ATLAS-EX.} = 1.39^{+0.17}_{-0.16},
\label{eq:sigstrenth_ttW_atlas}
\end{equation}
which corresponds to a {$2.4\sigma$} discrepancy.
Importantly, the prediction is built from an established~\cite{deFlorian:2016spz} reference cross section of $\sigma^{\rm ref.}_{\ttW} = 601\fb^{+13\%}_{-12\%}$ that includes leading QCD and EW corrections, accounts for additional EW corrections~\cite{Frederix:2017wme}  through a scaling factor  {$K_{\rm sub-NLO-EW}=1.09$}, but  also  includes a scaling factor {$K^{\rm est.}_{\rm NNLO-QCD}=1.11$} for contributions at  NNLO in QCD.

While seemingly innocuous, the estimate of $K^{\rm est.}_{\rm NNLO-QCD}$ is based on the observation \cite{Alwall:2014hca} that the $pp\to\ttWj$ process exhibits a large, $\mathcal{O}(40\%)$ correction at  NLO in QCD  for a specific set of inputs. However, neither Ref.~\cite{Alwall:2014hca} nor follow-up work~\cite{Maltoni:2015ena} evaluate the $\ttW$ rate beyond $\mathcal{O}(\alpha_s^3 \alpha)$. Therefore, owing to the uncertainty in $K^{\rm est.}_{\rm NNLO-QCD}$, we turn our focus to the roles of the $\ttWj$ and $\ttWjj$ sub-processes in inclusive $\ttW$ production.

%%%%%%%%%%%%%%%%%%%%%%%%%%%%%%%%%%%%%%%%%%%%%%%%%
\subsection{The \ttWj and \ttWjj processes}\label{sec:inclusive_ttwjx}

\begin{figure}
    \centering
    \includegraphics[width=.9\columnwidth]{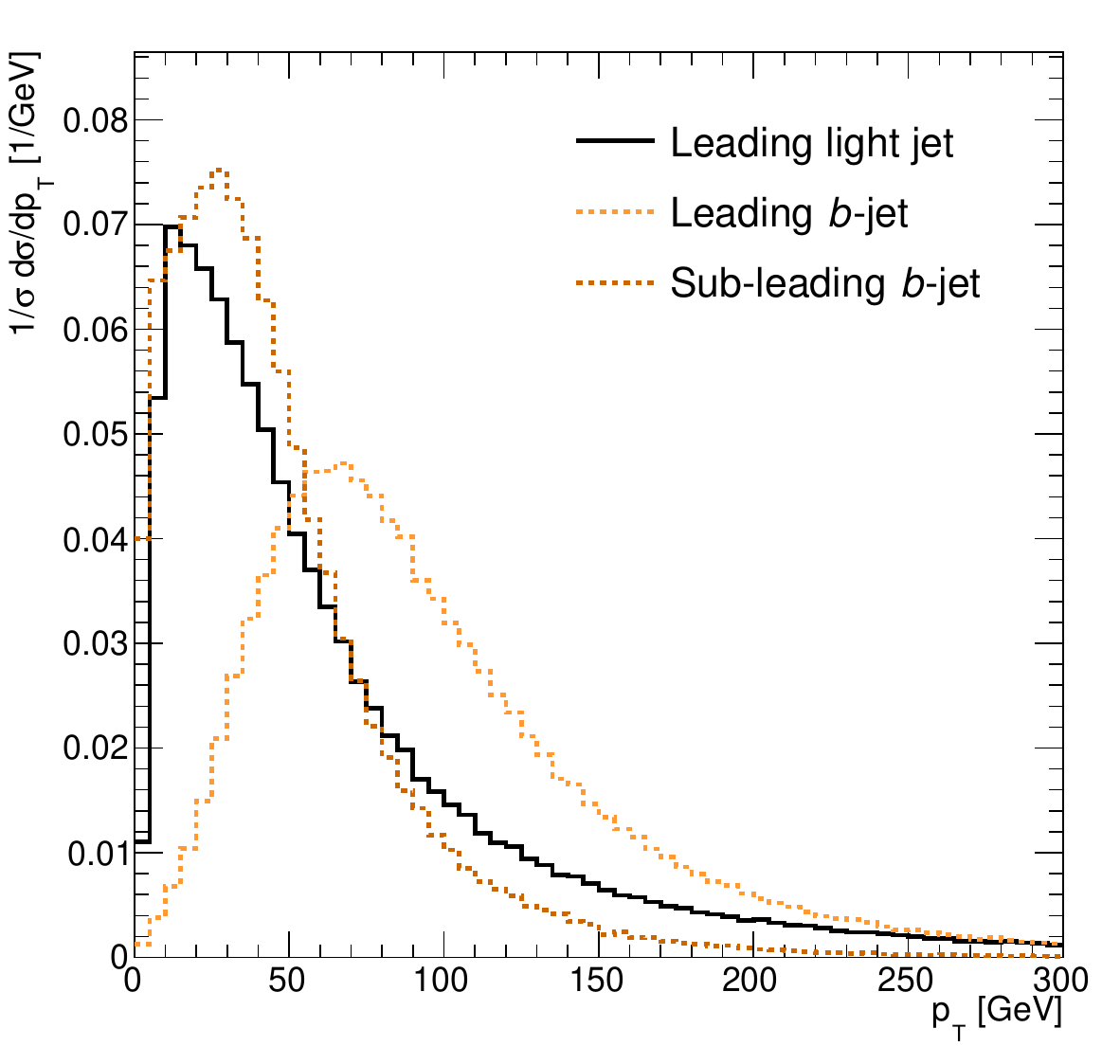}
    \caption{Normalized $p_T$ distributions of the leading  $b$-jet (light dash), sub-leading  $b$-jet (dark dash), and leading light jet (solid) in the $pp\to \ttW\to 3\ell+X$ process at NLO in QCD with PS matching.}
    \label{fig:ttWAnatomy_pT_jets_ttW}
\end{figure}

To investigate the $pp\to\ttWj$ and $pp\to\ttWjj$  processes, we argue first for definitions of these channels that ensure their matrix elements (MEs) are perturbative in the CSS sense~\cite{Collins:1984kg,Collins:2011zzd}, i.e., are absent of large collinear logarithms. Such logarithms originate from real radiation that go soft or collinear and require sufficiently stringent transverse momentum cuts $(p_T^{j~\min})$ to render MEs physical.
For a $(t\overline{t}W)$-system invariant mass of $M_{t\overline{t}W}$, these contributions cause cross sections to scale as
\begin{equation}
\sigma(pp\to \ttWpm+nj)\sim \sum_{k=n} \alpha_s^{k}\left(p_T^{j~\min}\right)\log^{2k-1}\left(\frac{M^2_{t\overline{t}W}}{p_T^{j~\min~2}}\right).
\label{eq:css_expansion}
\end{equation} 
For too small $p_T^{j~\min}$ one enters the Sudakov regime where $\log$ factors exceed $1/\alpha_s$ factors and $k_T$-resummation is needed.

To establish a sufficiently ``safe'' $p_T^{j~\min}$, we consider at $\sqrt{s}=13\TeV$ the  $pp\to\ttW$ process at NLO in QCD, i.e., up to $\mathcal{O}(\alpha_s^3 \alpha)$, and the LO decay to charged leptons,
\begin{equation}
pp  \to \ttWpm \to 3W ~b\overline{b} \to 3\ell ~3\nu ~b\overline{b}, \quad \ell\in\{e,\mu\}.
\end{equation}
Here and throughout we choose the dynamic scale scheme of equation~\ref{eq:scale} following studies~\cite{Berger:2009ep,Campbell:2017hsr}  of the $pp\to W+nj$ process. There, schemes not reflective of kinematic scales were shown to lead to negative cross sections at NLO. For the inclusive $\ttW$ process at NLO (LO) in  QCD, we obtain
\begin{equation}
    \sigma_{\ttWpm}^{\rm NLO-QCD~(LO)} = 594\fb~^{+11\%}_{-10\%}~^{+2.0\%}_{-2.0\%} ~ (378\fb~^{+24\%}_{-18\%}~^{+2.2\%}_{-2.2\%}),
    \label{eq:xsec_nlo}
\end{equation}
as our baseline NLO (LO) cross sections. The first and second uncertainty reflect scale and PDF dependence, respectively.

After parton showering, anti-$k_T$ clustering  $(R=0.4)$, and overlap removal between leptons and jets, we plot in figure~\ref{fig:ttWAnatomy_pT_jets_ttW} the normalized $p_T$ distribution of the leading light jet (solid). As a check of our computational setup, we also plot the leading (light dash) and sub-leading (dark dash) $b$-jet.  The $b$-jet distributions reflect the characteristic momentum $p_T\sim m_t(1-M_W^2/m_t^2)/2\sim65\GeV-70\GeV$, modulo recoils against  $W$ and light jets.

Since the $\ttW$ system is fully decayed to leptons, the leading light jet is an $\mathcal{O}(\alpha_s^3 \alpha)$ contribution that originates, in the MC@NLO formalism~\cite{Frixione:2002ik},   from (i) the tree-level $\ttWj$ ME at ``large'' $(p_T^j/M_{t\overline{t}W})$ or (ii) PS corrections to the one-loop-improved $\ttW$ ME at ``small'' $(p_T^j/M_{t\overline{t}W})$. 
Using the procedure in Ref.~\cite{Degrande:2016aje}, which generalizes an analogous procedure in Ref.~\cite{Collins:1984kg}, we  estimate that the transition between the two domains occurs at around $p_T^{\rm safe}\sim30\GeV$ for $M_{t\overline{t}W}\sim425\GeV-475\GeV$. Notably, this estimate neglects $\beta_0$ factors in $\alpha_s$ running. Accounting for this we obtain instead  $p_T^{\rm safe}\sim95\GeV-100\GeV$.  In comparison to figure~\ref{fig:ttWAnatomy_pT_jets_ttW} one sees that the transition between the two domains occurs somewhere\footnote{For comparison, the Sudakov peak for $W$ production occurs at $p_T=3\GeV-4\GeV$, is estimated to transition to fixed order MEs at $p_T\sim7\GeV-8\GeV$ \cite{Collins:1984kg}, and does so at $p_T\sim5\GeV-10\GeV$ \cite{Abbott:2000xv}.}  between the Sudakov peak at $p_T\sim15\GeV-20\GeV$ and $p_T\sim75\GeV-100\GeV$, with an inflection point at  $p_T\sim50\GeV$. This is also roughly the geometric mean of the estimated $p_T^{\rm safe}$.  This suggests that  regulators  below  $p_T^{j~\min}\sim30\GeV-50\GeV$ will lead  to unphysical  MEs, intermediate cutoffs of $p_T^{j~\min}\sim50\GeV-75\GeV$ can arguably stabilize MEs, and that  cutoffs of $p_T^{j~\min}\gtrsim100\GeV$ are fine.

\begin{table}[!t]
\begin{center}
%\resizebox{\textwidth}{!}{
\resizebox{\columnwidth}{!}{
\begin{tabular}{c r c c l c c}
\hline\hline
Order & $p_T^{j~\min}$ & $\sigma$ [fb]  & $\pm\delta_{\mu_f,\mu_r}$ & $\pm\delta_{\rm PDF}$ & $K_{\rm QCD}$ & $\Delta\sigma$ [fb] \\
\hline
LO  	& 30\GeV    	&   227 & $^{+40\%}_{-27\%}$ & $^{+1.3\%}_{-1.3\%}$ & \dots & \dots \\
	& 40\GeV    	&   191 & $^{+41\%}_{-27\%}$ & $^{+1.3\%}_{-1.3\%}$ & \dots & \dots \\
    	& 50\GeV    	&   164 & $^{+41\%}_{-27\%}$ & $^{+1.2\%}_{-1.2\%}$ & \dots & \dots \\
	& 75\GeV		&  122  & $^{+42\%}_{-28\%}$ & $^{+1.1\%}_{-1.1\%}$ & \dots & \dots \\
    	& 100\GeV   	&  93.5 & $^{+42\%}_{-28\%}$ & $^{+1.1\%}_{-1.1\%}$ & \dots & \dots \\
	& 125\GeV	&  74.5 & $^{+42\%}_{-28\%}$ & $^{+1.3\%}_{-1.3\%}$ & \dots & \dots \\
    	& 150\GeV   	&  59.8 & $^{+43\%}_{-28\%}$ & $^{+1.1\%}_{-1.1\%}$ & \dots & \dots \\
\hline
NLO	& 30\GeV    	& 	351	& $^{+12\%}_{-14\%}$ & $^{+1.2\%}_{-1.2\%}$ & 1.55 & 124 \\
	& 40\GeV    	& 	303	& $^{+13\%}_{-15\%}$ & $^{+1.1\%}_{-1.1\%}$ & 1.59 & 112 \\
	& 50\GeV    	& 	267	& $^{+14\%}_{-15\%}$ & $^{+1.0\%}_{-1.0\%}$ & 1.62 & 103 \\
	& 75\GeV		&	205	& $^{+16\%}_{-16\%}$ & $^{+1.0\%}_{-1.0\%}$ & 1.68 & 83.0 \\
	& 100\GeV	&	159	& $^{+16\%}_{-16\%}$ & $^{+0.9\%}_{-0.9\%}$ & 1.70 & 65.7 \\
	& 125\GeV	&	129	& $^{+17\%}_{-17\%}$ & $^{+0.9\%}_{-0.9\%}$ & 1.73 & 54.7 \\
	& 150\GeV	& 	104	& $^{+17\%}_{-17\%}$ & $^{+0.9\%}_{-0.9\%}$ & 1.73 & 43.9\\
\hline\hline
\end{tabular}
} %% resize
\caption{
Total cross sections [fb] at $\sqrt{s}=13\TeV$ of the $pp\to\ttWpmj+X$ process at LO and NLO in QCD, with scale and PDF uncertainties [\%] for representative  jet $p_T$ thresholds $(p_T^{j~\min})$  with $\vert\eta^j\vert<4.0$.
Also shown are the $K$-factors and differences between NLO and LO rates.
} \label{tab:normttW1j}
\end{center}
\end{table}

To check this, for representative $p_T^{j~\min}$ with $\vert\eta^j\vert<4.0$,
we compute and list in table~\ref{tab:normttW1j} cross sections at  LO and NLO in QCD, i.e., up to $\mathcal{O}(\alpha_s^3\alpha)$ and $\mathcal{O}(\alpha_s^4\alpha)$,  for the process
\begin{equation}
    p p \to \ttWpmj,
\end{equation}
with scale and PDF uncertainties, the QCD $K$-factor
\begin{equation}
    K_{\rm  QCD} ~\equiv~ \sigma^{\rm N^{k+1}LO}  ~/~ \sigma^{\rm N^kO},
    \label{eq:kFactQCD}
\end{equation}
and the difference between cross sections at NLO and LO
\begin{equation}
    \Delta\sigma \equiv \sigma_{\ttWj}^{\rm NLO} - \sigma_{\ttWj}^{\rm LO},
    \label{eq:xsecDiff}
\end{equation}
which quantifies $\mathcal{O}(\alpha_s^4\alpha)$ contributions. For $p_T^{j~\min}=30\GeV-150\GeV$, NLO rates span $\sigma_{\ttWj}^{\rm NLO}\sim100\fb-350\fb$, in agreement with Refs. \cite{Maltoni:2015ena,Alwall:2014hca} when assuming their theoretical inputs.
We report that scale uncertainties are uniform across $p_T^{j  \min}$  and reduce from {$30\%-40\%$} at LO to {$15\%$} at NLO. This suggests perturbative stability for $p_T^{j~\min}\gtrsim50\GeV-75\GeV$. 

As in the inclusive $\ttW$ rate at LO and NLO in QCD, predictions for $\ttWj$ are acutely sensitive to choices of scale, and ultimately to the running of $\alpha_s(\mu_r)$.
For example: the authors of Ref.~\cite{Maltoni:2015ena} argue in favor of scale scheme that takes the geometric mean of particles' transverse energies. In practice, this leads to $\mu_f,\mu_r$ that are about $30\%-40\%$ smaller than the scheme we use (see equation \ref{eq:scale}), which instead sums particles' transverse energies. While one would na\"ively expect only a minor shift in rate normalization, the authors of Ref.~\cite{Maltoni:2015ena} also employ one-loop (two-loop) running of $\alpha_s(\mu_r)$ in a LO (NLO) PDF for their LO (NLO) in QCD computations. As documented in section \ref{sec:mc}, we employ three-loop running in an NNLO PDF in all computations to make explicit the impact of ME corrections and avoid possible double counting of $\mathcal{O}(\alpha_s^2)$ contributions. This implies that the value of $\alpha_s(\mu_r)$ employed in Ref.~\cite{Maltoni:2015ena} is about $\alpha_s^{\rm 2-loop~(1-loop)}/\alpha_s^{\rm 3-loop}\sim 8\%~(25\%)$ larger for their NLO~(LO) calculation than for the analogus computation here. Remarkably, for three powers of $\alpha_s$ this compounds to a normalization shift in the total cross section of about $28\%~(96\%)$. This accounts for most differences between the $\ttWj$ rates reported in Ref.~\cite{Maltoni:2015ena} and in table~\ref{tab:normttW1j} for $p_T^j=100\GeV$. In comparison to the results reported Ref.~\cite{Alwall:2014hca}, which uses the same scale choices as we do, the NLO (LO) rate at $p_T^j=30\GeV$ that we report in table~\ref{tab:normttW1j} is about $3\%$ larger ($3\%$ smaller) and again is due to differences in using NNLO and NLO PDF sets. The superficial differences between the rates reported in Refs. \cite{Maltoni:2015ena,Alwall:2014hca}  underscore the impact of higher-order QCD corrections and the difficulty in ascribing a theoretical uncertainty.

\begin{table}[!t]
\begin{center}
%\resizebox{\textwidth}{!}{
\resizebox{\columnwidth}{!}{
\begin{tabular}{c c c c l l l}
\hline\hline
\multicolumn{6}{c}{$i ~j ~\to ~t ~\overline{t} ~W^\pm ~k ~l$} \\
\hline
$(i,j)$  & $(k,l)$ & $p_T^{j_1~\min}$& $p_T^{j_2~\min}$ & $\sigma$ [fb]  & $\pm\delta_{\mu_f,\mu_r}$ & $\pm\delta_{\rm PDF}$ \\
\hline
All             & All       & 75\GeV  & 75\GeV   	& 34.7~(100\%)   & $^{+57\%}_{-34\%}$    &  $^{+1.1\%}_{-1.1\%}$ \\
$(g,\mathcal{Q})$ & $(g,\mathcal{Q})$					&  &	& 23.7~(68\%)  & & \\                
$(\mathcal{Q},\mathcal{Q})$ & $(\mathcal{Q},\mathcal{Q})$	&  &	& 6.99~(20\%)  &   &\\                            
$(\mathcal{Q},\mathcal{Q})$ & $(g,g)$					&  &	& 3.63~(10\%)  &   &\\
$(g,g)$         & $(q,\overline{q})$						&  &	& 0.437~(1.3\%) & & \\
\hline
All     & All       & 100\GeV  & 75\GeV   		& 33.1~(100\%)   & $^{+57\%}_{-34\%}$ & $^{+1.0\%}_{-1.0\%}$ \\
$(g,\mathcal{Q})$ & $(g,\mathcal{Q})$                       			&  &   & 22.6~(68 \%)  & & \\                
$(\mathcal{Q},\mathcal{Q})$ & $(\mathcal{Q},\mathcal{Q})$   	&  &   & 6.78~(20\%)  &   & \\                       
$(\mathcal{Q},\mathcal{Q})$ & $(g,g)$                       			&  &   & 3.28~(9.9\%)  &   &\\
$(g,g)$                     & $(q,\overline{q})$            				&  &   & 0.409~(1.2\%) & & \\
\hline
All             & All       & 100\GeV  & 100\GeV   	& 21.2~(100\%)   & $^{+57\%}_{-34\%}$ & $^{+1.1\%}_{-1.1\%}$ \\
$(g,\mathcal{Q})$ & $(g,\mathcal{Q})$   &  &   & 14.3~(67\%)  & & \\                
$(\mathcal{Q},\mathcal{Q})$ & $(\mathcal{Q},\mathcal{Q})$   &  &   & 4.91~(23\%)  &   & \\                       
$(\mathcal{Q},\mathcal{Q})$ & $(g,g)$   &  &   & 1.75~~(8\%)  &   &\\
$(g,g)$                     & $(q,\overline{q})$   &  &   & 2.58~~(1\%) & & \\
\hline
$(g,q_V)$ & $(g,q_V)$   & 75\GeV	& 75\GeV		& 20.1~(58\%)    & $^{+58\%}_{-35\%}$    &  $^{+2.3\%}_{-2.3\%}$ \\
$(g,q_V)$ & $(g,q_V)$   & 100\GeV	& 75\GeV		& 19.3~(58\%)    & $^{+58\%}_{-35\%}$    &  $^{+2.3\%}_{-2.3\%}$ \\
$(g,q_V)$ & $(g,q_V)$   &  100\GeV	& 100\GeV	& 12.2~(58\%)  &  $^{+59\%}_{-35\%}$    & $^{+2.4\%}_{-2.4\%}$ \\
\hline\hline
\end{tabular}
} %% resize
\caption{
Total cross sections [fb] at $\sqrt{s}=13\TeV$ for the $pp\to\ttWpmjj$ process at LO, with scale and PDF uncertainties [\%], for representative  $p_T^{j_k~\min}$  with $\vert\eta^j\vert<4.0$. Also shown is the decomposition according to partonic channel, for $q_V\in\{u,d\}$, $q\in\{u,d,c,s\}$, and $\mathcal{Q}\in\{q,\overline{q}\}$.
} \label{tab:normttW2j}
\end{center}
\end{table}

Focusing on  table~\ref{tab:normttW1j}, two notable observations can be drawn from these rates. First is the size of the pure $\mathcal{O}(\alpha_s^4\alpha)$ contributions. For $p_T^{j~\min}=75\GeV-150\GeV$, these are positive and span about {$\Delta\sigma\sim 45\fb-85\fb$}. In comparison to the baseline $\ttW$ cross section in equation~\ref{eq:xsec_nlo}, we find  that $\Delta\sigma$  is  {$10\%-20\%$} of the baseline rate at LO and {$7\%-14\%$} at NLO for our range of $p_T^{j~\min}$. Specifically for $p_T^{j \min}=75\GeV-125\GeV$, we find that $\Delta\sigma$ alone is about {$20\%-30\%$} of the discrepancy reported in equation \ref{eq:sigstrenth_ttW_atlas}. The largeness of $\Delta\sigma$ is despite the fact that it constitutes an $\mathcal{O}(\alpha_s^2)$ correction to inclusive $\ttW$ production, which would otherwise suggest that $\Delta\sigma$ is at most $\mathcal{O}(1-10\%)$.  

Importantly, $\Delta\sigma$ encapsulates corrections to $\ttWj$ for a leading jet $j_1$ that is well-defined in the CSS sense; it does not reflect corrections for when $j_1$ is in the soft-wide angle limit. Extrapolating from table \ref{tab:normttW1j} hints that such corrections remain positive and that the $K^{\rm est.}_{\rm NNLO-QCD}$ in the previous section underestimates NNLO in QCD corrections by at least  a factor of two. Isolating this, however, is complicated by the phase space region where $j_1$ is hard but collinear to the beam line. Such regions cancel against negative-valued PDF/collinear counter terms at NNLO.

The second observation is that QCD corrections to the $\ttWj$ process are large, with $K$-factors ranging {$K_{\rm QCD}\sim1.7$ for $p_T^{j \min} = 75\GeV-150\GeV$}, and with the largest (smallest) $p_T^{j~\min}$ exhibiting the largest (smallest) corrections. Like \ttW, it is possible that new partonic channels associated with the $\ttWjj$ sub-process at $\mathcal{O}(\alpha_s^4\alpha)$ drive these increases. 

To check this we consider at LO, i.e., $\mathcal{O}(\alpha_s^4\alpha)$, the channel
\begin{equation}
    p p \to \ttWpmjj,
\end{equation}
and list in  table~\ref{tab:normttW2j} the $\sqrt{s}=13\TeV$ cross sections [fb] with uncertainties [\%], for benchmark $p_T^{j_k~\min}$  on the leading $(j_1)$ and sub-leading $(j_2)$ jets, and with $\vert\eta^j\vert<4.0$. Also shown is the decomposition according to initial/final-state partons.

For $p_T^{j_k \min}=75\GeV, 100\GeV$, we find that rates span {$\sigma_{\ttWjj}^{\rm LO}\sim 20\fb-35\fb$, with up to $60\%$ scale uncertainty and about a $1\%$ PDF uncertainty.}
This translates to about $5\%-9\% ~(3\%-6\%)$ of the baseline $\ttW$ rate at (N)LO.
  For all cases, quark-gluon scattering $(\mathcal{Q},g)$  accounts for about {$70\%$} of  the total $\ttWjj$ rate, whereas quark-quark scattering $(\mathcal{Q},\mathcal{Q})$ contributes {$20\%$}. This effectively rules out enhancements to $\ttW$ production at NNLO from valence-valence scattering. Instead, we find evidence of a large gluon-valence component, with  {$60\%$} of the  $\ttWjj$ rate being due to gluon-up/down scattering $(g,q_V)$.

Returning to $\Delta\sigma$ in equation \ref{eq:xsecDiff}, we recall that it encapsulates the $\mathcal{O}(\alpha_s^4\alpha)$ parts of $\ttWj$ production at NLO in QCD. Given then the LO $\ttWjj$ rate, we can estimate the net impact of the soft-real, virtual, and counter-term corrections (SR+V+CT) to  $\ttWj$  by taking the difference of the two:
\begin{align}
    \Delta\sigma^{\rm SR+V+CT}_{\ttWj}&(p_T^{j_1 \min},p_T^{j_2 \min}) \nonumber\\
    & \equiv  \Delta\sigma_{\ttWj}(p_T^{j_1 \min}) - \sigma_{\ttWjj}^{\rm LO}(p_T^{j_1 \min},p_T^{j_2 \min}).
\end{align}
After doing so, we find that the absolute [fb] (relative [\%]) unresolved corrections to the $\ttWj$ process at NLO are 
\begin{subequations}
\begin{align}
 \Delta\sigma^{\rm SR+V+CT}_{\ttWj}(75\GeV,75\GeV) &= +48.3\fb~(+58\%),		\label{eq:virt_loPT_loPT}\\
  \Delta\sigma^{\rm SR+V+CT}_{\ttWj}(100\GeV,75\GeV) &= +32.6\fb~(+50\%),		\label{eq:virt_hiPI_loPT}\\
 \Delta\sigma^{\rm SR+V+CT}_{\ttWj}(100\GeV,100\GeV) &= +44.5\fb~(+68\%).	\label{eq:virt_hiPT_hiPT}
\end{align}
\label{eq:virt_xPT_xPT}
\end{subequations}
For $p_T^{j_1 \min}=p_T^{j_2 \min}=75\GeV-100\GeV$, we find that SR+V+CT corrections constitute  about 
{$60\%-70\%$} of  $\Delta\sigma$. This suggests  that the $\mathcal{O}(\alpha_s^4\alpha)$ contributions at $p_T^{j_k \min}=75 (100)\GeV$ are dominated by unresolved radiation. For  mixed $p_T^{j_k \min}$,  SR+V+CT corrections dip to {$50\%$}, demonstrating an interplay between resolved and unresolved radiation. (This includes the role of the hierarchy $p_T^{j_1 \min} \gg p_T^{j_2 \min}$, which introduces logarithmic structures not captured in equation \ref{eq:css_expansion}~\cite{Dasgupta:2001sh,Forshaw:2006fk}.)  Importantly, the interplay in equation \ref{eq:virt_xPT_xPT} highlights that adding $\Delta\sigma_{\ttWj}$ at low $p_T^{j \min}$ to inclusive $\ttW$  should be accompanied by a reweighting / subtraction scheme that systematically protects against double counting of low-$p_T$ radiation.

In summary,  we report the existence of partonic configurations at $\mathcal{O}(\alpha_s^4\alpha)$, i.e., pure $\mathcal{O}(\alpha_s^2)$ corrections to inclusive $\ttW$ production, with cross sections in well-defined phase space regions that greatly exceed estimates by standard scale variation at NLO in QCD. By one measure (table~\ref{tab:normttW1j}) a subset of these corrections span at least {$\Delta \sigma \sim 65\fb-85\fb$},  or about {$10\%-14\%$} of the inclusive rate at NLO in QCD,
and follows from a mixture of gluon-valence scattering in $\ttWjj$ at LO (table~\ref{tab:normttW2j}) and unresolved radiation in $\ttWj$ at NLO (equation \ref{eq:virt_xPT_xPT}). While alone accounting for {$20\%-30\%$} of the discrepancy in measured the $\ttW$  rate at $\sqrt{s}=13\TeV$ (see, for example, equation \ref{eq:sigstrenth_ttW_atlas}), other contributions at $\mathcal{O}(\alpha_s^4\alpha)$, such as  two emissions of soft, wide-angle legs and one soft leg at one-loop, which are arguably positive, are not included. Importantly, resummed results at NNLL do not suggest large cancelations against pure two-loop diagrams. Guided by this, we turn to the impact of combining the $\ttW$ and $\ttWj$ processes using NLO multi-jet matching techniques.

%%%%%%%%%%%%%%%%%%%%%%%%%%%%%%%%%%%%%%%%%%%%%%%%%
\subsection{Inclusive production with NLO multi-jet matching}\label{sec:inclusive_fxfx}
 
As a complementary estimate of the $\mathcal{O}(\alpha_s^4\alpha)$ contributions to $\ttW$ production that stem from the  $\ttW+nj$ sub-channels we employ the FxFx NLO multi-jet matching \cite{Frederix:2012ps}. In short, FxFx is an established \cite{TheATLAScollaboration:2016sdf,Khachatryan:2016mnb,Aaboud:2018xdt,Sirunyan:2018cpw}, non-unitarity \cite{Frederix:2012ps,Frederix:2015eii} procedure within the MC@NLO formalism for promoting jet observables at LO+LL  to NLO+LL through CKKW-like~\cite{Catani:2001cc} reweighting. In particular, hard, wide-angle emissions are included through exact MEs at one-loop and double counting is avoided by Sudakov reweighting. As such, cross sections at NLO are augmented with terms that are $\mathcal{O}(\alpha_s^2)$ or higher. 

The cost of this improvement is the introduction of a merging scale $(Q_{\rm cut}^{\rm FxFx})$ akin to those at LO. To set $Q_{\rm cut}^{\rm FxFx}$ we follow  Refs.~\cite{Alwall:2014hca,Frederix:2012ps,Frederix:2015eii}, which call for $Q_{\rm cut}^{\rm FxFx} > 2 p_T^{j \min}$. In principle, FxFx merging is independent of $p_T^{j \min}$ and ultra-low $p_T^{j \min}$ choices simply lead to large event-veto rates and therefore poorer Monte Carlo efficiency. Formally, however, the Sudakov-reweighting in FxFx only cancels the collinear logarithms that are shared by MEs and the PS; for sufficiently small $p_T^{j \min}$, mis-cancellations of soft logarithms can technically spoil perturbative convergence. Therefore, as a jet $p_T$ threshold is needed to regulate Born-level $t\overline{t}W+nj$ MEs, we also require that $p_T^{j_k \min}$  is not too low in the CSS sense and  that $\vert\eta^j\vert<4.0$. We match up to the first jet multiplicity (denoted as FxFx1j) and  set as our baseline configuration  
\begin{equation}
{(p_T^{j \min},Q_{\rm cut}^{\rm FxFx})=(50\GeV, 110\GeV)}.
\label{eq:fxfx_setup}
\end{equation}

At this order we obtain as the inclusive $\ttW$ cross section
\begin{equation}
    {\sigma_{\ttWpm}^{\rm FxFx1j} = 
    655\fb~^{+12\%}_{-12\%}~^{+1.6\%}_{-1.6\%}},
        \label{eq:xsec_fxfx}
\end{equation}
where the uncertainties reflect the scale and PDF dependence, respectively.
We report that corrections at this order increase the baseline NLO rate in equation \ref{eq:xsec_nlo} by about {$10\%$} with the size of uncertainties remaining essentially the same. This is in agreement with the estimated~\cite{ATLAS:2019nvo} $K^{\rm est.}_{\rm NNLO-QCD}=1.11$, and hence is at odds with  fixed-order estimates in table \ref{tab:normttW1j}, which suggest that such $\mathcal{O}(\alpha_s^4\alpha)$ contributions are larger.

To be more precise, an important difference between the fixed-order estimates of $\mathcal{O}(\alpha_s^4\alpha)$ contributions and the FxFx rate is the treatment of soft corrections to the $\ttWj$ sub-process. As described above, the FxFx matching scheme does not fully account for soft logarithmic corrections in its Sudakov reweighting. It only accounts for collinear logarithms shared by the ME and the PS. If both estimates are to be trusted up to their formal accuracies, then it is possible that  soft corrections to the $\ttWj$ sub-process are indeed sizable. This discrepancy suggests a need for the full NNLO in QCD description of inclusive $\ttW$ production at the LHC.

Accounting now for the EW corrections in equation \ref{eq:kFactEW}~\cite{Frederix:2017wme} and assuming the FxFx uncertainties above, we obtain
\begin{align}
    \sigma_{\ttWpm}^{\rm FxFx1j+EW} \equiv \sigma_{\ttWpm}^{\rm FxFx1j} + \delta\sigma_{\rm EW}^{\rm NLO} 
        = 690\fb~^{+12\%}_{-12\%}~^{+1.6\%}_{-1.6\%}.
        \label{eq:xsec_fxfx_ew}
\end{align}
We find that this rate is about {$5\%$} smaller than the prediction used in the ATLAS measurement of Ref. \cite{ATLAS:2019nvo}, and that the difference is mainly due to the scale and PDF choices in the baseline NLO in QCD rate.  (A  {$1\%$} difference follows from our FxFx correction being smaller than the estimated NNLO $K$-factor.) In principle, this revised cross section worsens slightly the discrepancy reported in equation \ref{eq:sigstrenth_ttW_atlas}.
For the pure FxFx and FxFx+EW cases, the  corresponding best-fit signal strengths are
\begin{align}
\hat{\lambda}_{\ttW}^{\rm FxFx1j} 		&= 1.54~^{+0.19}_{-0.18} \label{eq:sigstrenth_ttW_fxfx}, \\
\hat{\lambda}_{\ttW}^{\rm FxFx1j+EW} 	&= 1.46~^{+0.18}_{-0.17} \label{eq:sigstrenth_ttW_fxfx_ew},
\end{align}
and are consistent with SM expectations at {$2.7\sigma-3.0\sigma$}.

While NNLL threshold corrections can improve this picture, direct application of Ref. \cite{Kulesza:2018tqz} is hindered by the different scale choices that we assume. That said, taking a comparable correction of $K_{\rm NNLL-QCD}^{est.} = \sigma^{NLO+NNLL-QCD}/\sigma^{NLO-QCD} = 1.03$,  the associated SM rate and best-fit signal strength are:
\begin{align}
 \sigma_{\ttWpm}^{\rm FxFx1j+EW+NNLL}		&= 708\fb~^{+12\%}_{-12\%}~^{+1.6\%}_{-1.6\%},
									\label{eq:xsec_fxfx_ew_nnll}
\\
\hat{\lambda}_{\ttW}^{\rm FxFx1j+EW+NNLL}		&= 1.43~^{+0.17}_{-0.16},
									\label{eq:sigstrenth_ttW_fxfx_ew_nnll}
 \end{align}
where we again assume the FxFx uncertainties. With these estimated corrections the discrepancy stays at $2.7\sigma$.

To explore the uncertainty associated with our baseline $(p_T^{j \min},Q_{\rm cut}^{\rm FxFx})$, we report in the upper panel of table \ref{tab:normXSecFxFx} the inclusive $\ttW$ cross section  at $\sqrt{s}=13\TeV$ at LO, NLO in QCD, and with FxFx1j matching for various inputs. Also shown are the $\mu_f, \mu_r$ scale and PDF uncertainties, and the QCD $K$-factor as defined in  equation \ref{eq:kFactQCD}. We denote our benchmark rate by $\dagger$. 

For the inputs considered we observe that the NLO multi-jet matching rates span about  $\sigma^{\rm FxFxj1} \sim 600\fb-670\fb$. This is about $0.3\%-12\%$ larger than the baseline rate at NLO in QCD in equation \ref{eq:xsec_nlo}. Notably, the range of {$\Delta \sigma^{\rm FxFxj1} \sim 70\fb$} is much smaller than the differences in the $\ttWj$ rate at NLO, which span $\Delta \sigma \sim 250\fb$ (see  table \ref{tab:normttW1j}). Na\"ively, this indicates a sizable phase space overlap between the $\ttW$ and $\ttWj$ processes at NLO in QCD. However, this appears contrary to figure \ref{fig:ttWAnatomy_pT_jets_ttW}, which shows that the characteristic light jet scale in inclusive $\ttW$ is well below $p_T^j = 50\GeV$, and otherwise suggests a much milder  phase space overlap for $p_T^{j \min}\gtrsim 50\GeV-75\GeV$.

As expected~\cite{Frederix:2012ps,Frederix:2015eii}, we find that the range of FxFx1j predictions is driven by the dependence on the merging scale more than the jet $p_T$ threshold. In particular, for the largest $Q_{\rm cut}^{\rm FxFx}$ considered the FxFx1j rate reduces to the NLO rate and can be tied to an ``over suppression'' of the $\ttWj$ multiplicity~\cite{Hamilton:2012rf}. 
To quantify an uncertainty associated with $Q_{\rm cut}^{\rm FxFx}$, we consider the envelope spanned by all FxFx predictions. For $Q_{\rm cut}^{\rm FxFx} = 70\GeV-350\GeV$ and $p_T^{j \min}=30\GeV-150\GeV$, we report a variation of $\delta \sigma^{\rm FxFx1j}[Q_{\rm cut}^{\rm FxFx} ]/\sigma^{\rm FxFx1j}_{\rm baseline} = ~^{+2\%}_{-9\%}$.
To quantify an uncertainty associated with $p_T^{j \min}$, we fix the merging scale at 
\begin{equation}
Q_{\rm cut}^{\rm FxFx} = 110\GeV, ~150\GeV, ~250\GeV, ~350\GeV,
\end{equation}
and vary $p_T^{j \min}$. For all cases, we find that FxFx rates change only about $ \Delta \sigma^{\rm FxFxj1}[p_T^{j \min} ]  \sim 1\fb-10\fb$, or $0.1\%-1.5\%$.

%%%%%%%%%%%%%%%%%%%%%%%%%%%%%%%%%%%%%%%%%%%%%%%%%%%%%%%%%%%%%%%%%%%%%%%%%%%%%%
\section{Differential Production}\label{sec:differential}
%%%%%%%%%%%%%%%%%%%%%%%%%%%%%%%%%%%%%%%%%%%%%%%%%%%%%%%%%%%%%%%%%%%%%%%%%%%%%%

\begin{table}[!t]
\begin{center}
%\resizebox{\textwidth}{!}{
\resizebox{\columnwidth}{!}{
\begin{tabular}{l r r c l l c}
\hline\hline
Order & $p_T^{j~\min}$  & $Q_{\rm cut}^{\rm FxFx}$  & $\sigma$ [fb]  & $\pm\delta_{\mu_f,\mu_r}$ & $\pm\delta_{\rm PDF}$  & $K_{\rm QCD}$ \\
\hline
\multicolumn{6}{c}{\ttWpm (Inclusive)}\\
\hline
LO	& \dots		& \dots		&	378	& $^{+24\%}_{-18\%}$ & $^{+2.2\%}_{-2.2\%}$ & 1.00\\
NLO	& \dots		& \dots		&	594	& $^{+11\%}_{-10\%}$ & $^{+2.0\%}_{-2.0\%}$ & 1.57\\
\hline
FxFx1j & 30\GeV	& 70\GeV		&	668	& $^{+12\%}_{-13\%}$ & $^{+1.7\%}_{-1.7\%}$ & 1.12\\
FxFx1j & 30\GeV	& 110\GeV	&	656	& $^{+12\%}_{-13\%}$ & $^{+1.7\%}_{-1.7\%}$ & 1.11\\
FxFx1j & 30\GeV	& 150\GeV	&	634	& $^{+12\%}_{-13\%}$ & $^{+1.6\%}_{-1.6\%}$ & 1.07\\
FxFx1j & 30\GeV	& 250\GeV	&	616	& $^{+12\%}_{-13\%}$ & $^{+1.6\%}_{-1.6\%}$ & 1.04\\
FxFx1j & 30\GeV	& 350\GeV	&	596	& $^{+12\%}_{-12\%}$ & $^{+1.6\%}_{-1.6\%}$ & 1.00\\
FxFx1j & 40\GeV	& 90\GeV		&	664	& $^{+12\%}_{-12\%}$ & $^{+1.7\%}_{-1.7\%}$ & 1.12\\
FxFx1j & 40\GeV	& 110\GeV	&	655	& $^{+12\%}_{-12\%}$ & $^{+1.7\%}_{-1.7\%}$ & 1.10\\
FxFx1j$^\dagger$ & 50\GeV	& 110\GeV	&	655	& $^{+12\%}_{-12\%}$ & $^{+1.6\%}_{-1.6\%}$ & 1.10\\
FxFx1j & 50\GeV	& 150\GeV	&	644	& $^{+12\%}_{-13\%}$ & $^{+1.6\%}_{-1.6\%}$ & 1.08\\
FxFx1j & 50\GeV	& 250\GeV	&	622	& $^{+13\%}_{-13\%}$ & $^{+1.7\%}_{-1.7\%}$ & 1.05\\
FxFx1j & 50\GeV	& 350\GeV	&	602	& $^{+12\%}_{-13\%}$ & $^{+1.7\%}_{-1.7\%}$ & 1.01\\
FxFx1j & 100\GeV	& 250\GeV	&	615	& $^{+12\%}_{-12\%}$ & $^{+1.7\%}_{-1.7\%}$ & 1.03\\
FxFx1j & 100\GeV	& 350\GeV	&	597	& $^{+12\%}_{-12\%}$ & $^{+1.8\%}_{-1.8\%}$ & 1.00\\  
FxFx1j & 150\GeV	& 350\GeV	&	597	& $^{+12\%}_{-12\%}$ & $^{+1.8\%}_{-1.8\%}$ & 1.00\\
\hline
\multicolumn{6}{c}{\ttZ  (Inclusive)}\\
\hline
LO	    &    \dots		& \dots		& 502	& $^{+30\%}_{-22\%}$ & $^{+1.2\%}_{-1.2\%}$ 		& 1.00\\
NLO	    & \dots		& \dots		& 744	& $^{+9.5\%}_{-11\%}$ & $^{+1.1\%}_{-1.1\%}$ 	& 1.48\\
\hline
FxFx1j 	& 40\GeV	& 90\GeV	& 822	& $^{+6.2\%}_{-12\%}$ & $^{+1.1\%}_{-1.1\%}$ 	& 1.10 \\
FxFx1j$^\dagger$ 	& 50\GeV	& 110\GeV	& 836	& $^{+8\%}_{-12\%}$ & $^{+1.1\%}_{-1.1\%}$ 	& 1.12 \\
FxFx1j 	& 100\GeV	& 250\GeV	& 820	& $^{+12\%}_{-13\%}$ & $^{+1.2\%}_{-1.2\%}$ 		& 1.10 \\
FxFx1j 	& 150\GeV	& 350\GeV	& 797	& $^{+11\%}_{-12\%}$ & $^{+1.2\%}_{-1.2\%}$ 		& 1.07 \\
\hline\hline
\end{tabular}
} %% resize
\caption{
Upper: At various perturbative orders and $(p_T^{j \min},Q_{\rm cut}^{\rm FxFx})$ assignments with $\vert\eta^j\vert<4.0$, the inclusive   $\ttW$ cross section [fb] at $\sqrt{s}=13\TeV$, along with residual scale and PDF uncertainties [\%].
Lower: Same for inclusive \ttZ.
Benchmark FxFx rates denoted by $^\dagger$. 
} \label{tab:normXSecFxFx}
\end{center}
\end{table}

\begin{figure*}[!t]
\begin{center}
\subfigure[]{\includegraphics[width=.9\columnwidth]{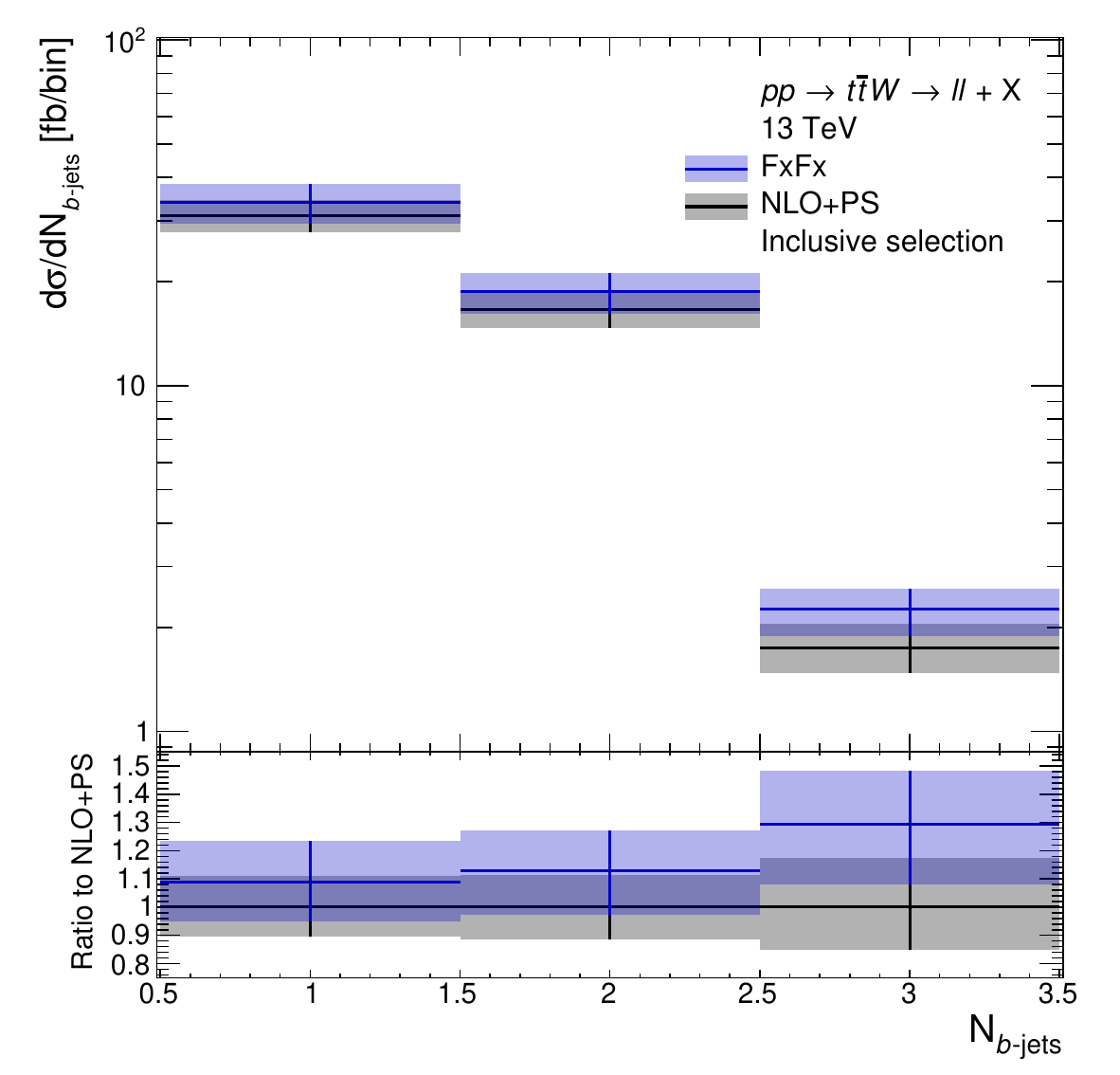}            \label{fig:ttWAnatomy_NrBJets_Inc}}
\subfigure[]{\includegraphics[width=.9\columnwidth]{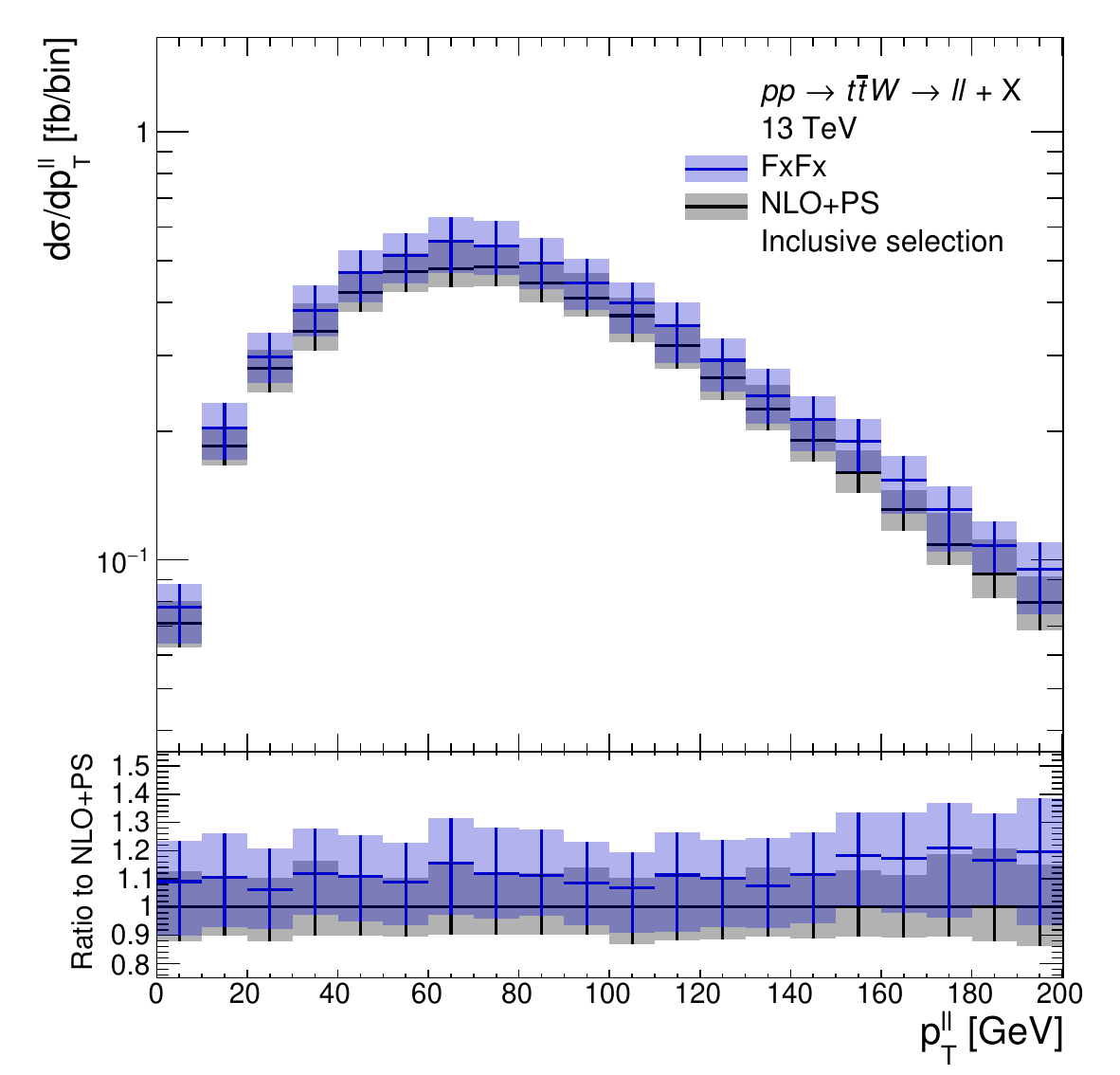}            \label{fig:ttWAnatomy_pT_ssll_Inc}}
\\
\subfigure[]{\includegraphics[width=.9\columnwidth]{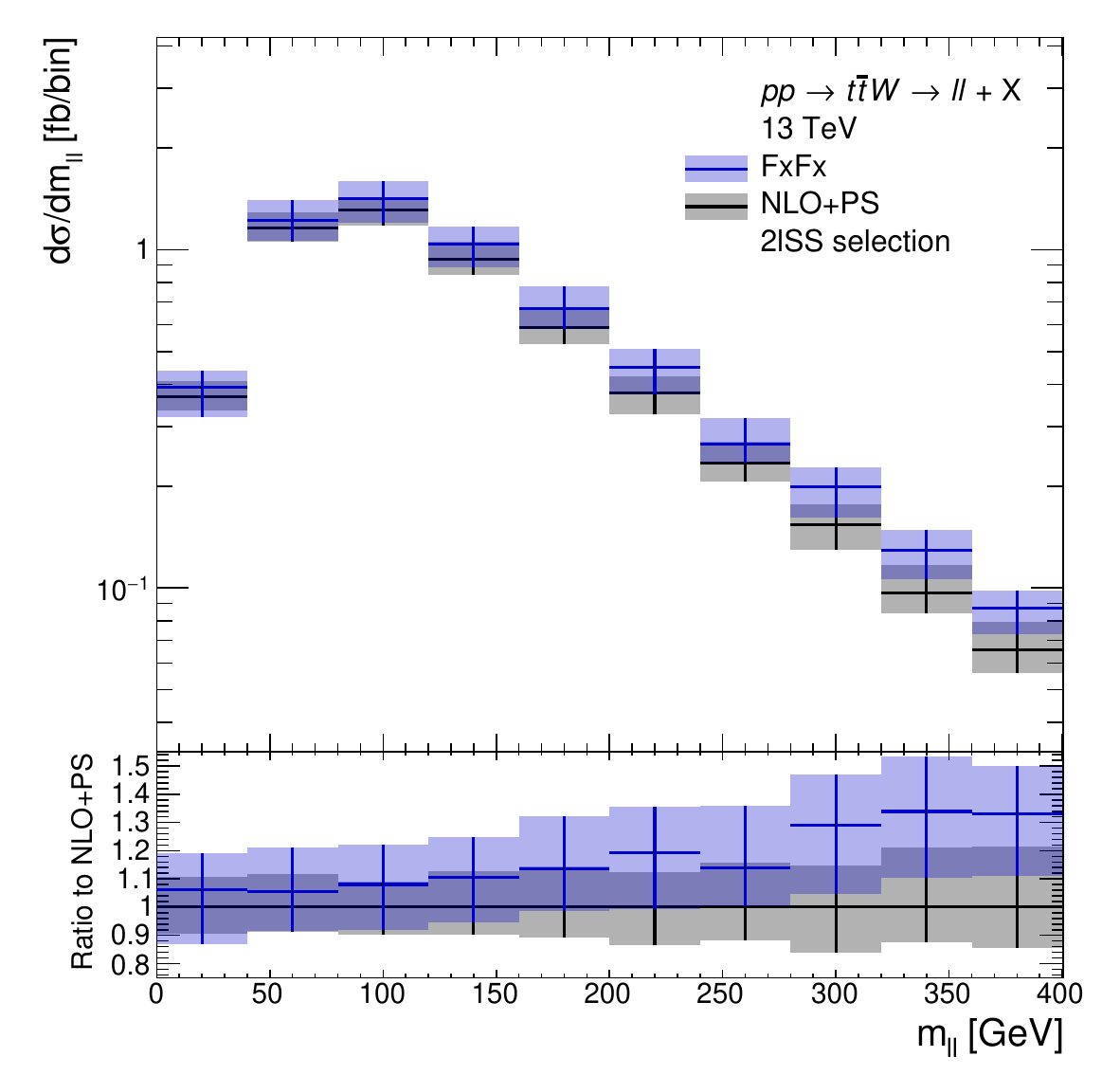}            \label{fig:ttWAnatomy_mX_ssll_2lX}}
\subfigure[]{\includegraphics[width=.9\columnwidth]{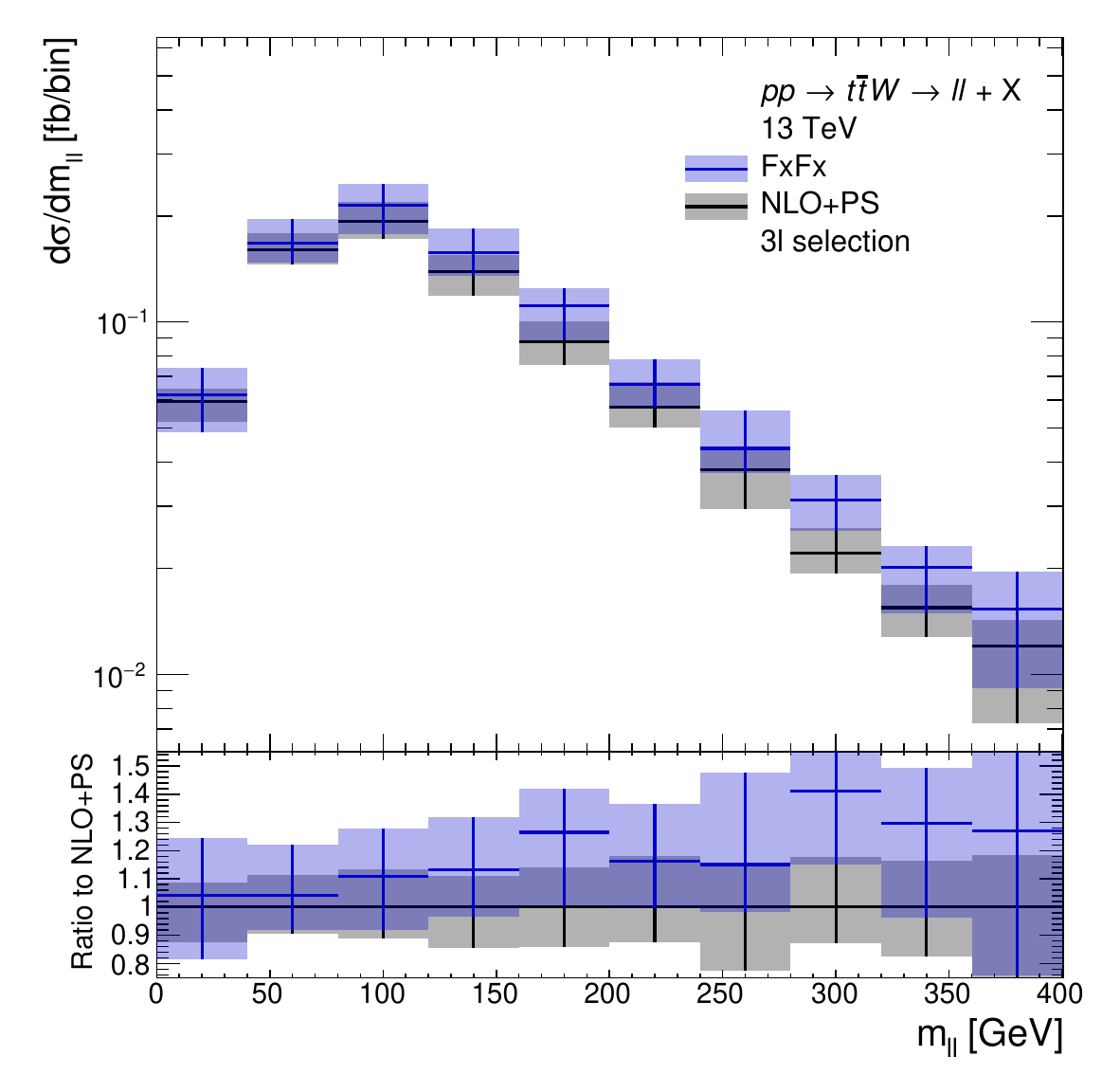}            \label{fig:ttWAnatomy_mX_ssll_3lX}}
\caption{
Upper: Differential cross sections with scale and PDF uncertainty envelopes at $\sqrt{s}=13\TeV$  of the $pp\to\ttW\to \ell^\pm_i \ell^\pm_j +X $ process at NLO in QCD with PS-matching (dark) and with FxFx matching (light), with respect to the
(a) $b$-jet multiplicity and (b) $p_T$ of the same-sign dilepton system in the inclusive selection category,
as well as the invariant mass of the dilepton system in the 
(c) \twolep and
(d) \threelep categories.
Lower: Ratio of FxFx and NLO+PS rates.
}
\label{fig:ttWAnatomy_dxsec}
\end{center}
\end{figure*}

Assuming that the enlarged $\ttW$ cross section measurements are solely due to missing QCD corrections, then NLO multi-jet matching cannot be the full picture. At the same time, differences in initial/finite-state radiation, associated production of heavy flavors, and relative enhancements by virtual radiation all impact particle kinematics. Hence, complementary to the total rate itself, kinematic distributions provide a means to test and understand the modeling of inclusive $\ttW$ production.

A comprehensive investigation into the impact of NLO multi-jet merging on particle kinematics is beyond our present scope and left to future work. That said, to at least build a qualitative picture, we take as a benchmark the ATLAS analysis~\cite{ATLAS:2019nvo}  associated with the signal strength in equation \ref{eq:sigstrenth_ttW_atlas} and consider the $\ttW$ decay mode,
\begin{equation}
pp  \to \ttWpm \to 3W ~b\overline{b} \to \ell^\pm_i \ell_j^\pm  + X, \quad \ell\in\{e,\mu\}.
\label{eq:differential_proc}
\end{equation}
Here the two same-sign $W$ bosons decay leptonically and the odd-sign $W$ boson decays inclusively.
Following closely the selection criteria of {table 3} in Ref.~\cite{ATLAS:2019nvo}, then after selection cuts and vetoes,  three signal categories are defined:
\begin{itemize}
\item The ``inclusive selection category'' is identified as two same-sign, high-$p_T$ charged leptons $\ell$,  two reconstructed light jets, and at least one $b$-tagged jet.
\item The ``two same-sign leptons'' (\twolep) category assumes category (i) but vetos events with three or more $\ell$.
\item The ``three leptons'' (\threelep) category again assumes (i) but requires exactly three $\ell$ with a net charge of $\pm1$.  
\end{itemize}

In  figure \ref{fig:ttWAnatomy_dxsec}, we plot the differential  cross sections  at $\sqrt{s}=13\TeV$ for representative observables and  signal categories at NLO in QCD with PS-matching (black) and FxFx-matching (blue).
In the insets, we plot the ratio of the FxFx and NLO+PS rates. Uncertainty bands are built from the envelope encapsulating the 27-point $\mu_f, \mu_f, \mu_s$  variation and $1\sigma$ PDF uncertainty.

Starting with figure \ref{fig:ttWAnatomy_NrBJets_Inc}, we show the $b$-jet multiplicity $(N_{b-jets})$ in the inclusive selection category. As anticipated from its larger cross section, we find that the normalization of the FxFx distribution is systematically larger than the NLO+PS one by at least {$10\%$}. More specifically, the bin-by-bin normalization grows to about   {$13\%$ for $N_{b-jets}=2$ and $30\%$ for $N_{b-jets}=3$,} and stems from the opening of $g^*\to b\overline{b}$ splitting in $q \overline{q} \to $\ttW$ g^*$ production. While not shown, we report a slight suppression (enhancement) of low (high) light jet  multiplicities relative to NLO+PS. Notably, an enhanced rate at high multiplicities for both heavy and light jets is slightly favored by data \cite{ATLAS:2019nvo,ATLAS:2020hrf}.

For the same signal category, we show in figure \ref{fig:ttWAnatomy_pT_ssll_Inc} the $p_T$ of the same-sign dilepton system $(p_T^{\ell\ell})$. Over the range $p_T^{\ell\ell}=0\GeV-200\GeV$, we observe a slowly increasing bin-by-bin shift in the FxFx normalization relative to the NLO+PS normalization. The increases range from about a  {$10\%$} enhancement to about  {$20\%$}. We attribute the growing FxFx rate with increasing $p_T^{\ell\ell}$ to the larger hadronic activity that the dilepton system recoils against, i.e., the positive contributions from the $\mathcal{O}(\alpha_s)$ corrections to the $\ttWj$ sub-channel.

Focusing now on more exclusive signal regions, we plot  in figures \ref{fig:ttWAnatomy_mX_ssll_2lX} and \ref{fig:ttWAnatomy_mX_ssll_3lX}, respectively, the invariant mass $(m_{\ell\ell})$ of the same-sign dilepton system for the \twolep and \threelep categories. For both categories, we observe a qualitatively similar but quantitatively stronger trend than found for $p_T^{\ell\ell}$. Numerically, enhancements grow from about {$10\%$} to just over  {$30\%$}, with the importance of NLO multi-jet matching increasing for larger $m_{\ell\ell}$. We argue that the  increases at larger $m_{\ell\ell}$ are due to additional final-state radiation off top quarks in the $\ttWj$ sub-channel at NLO. Such radiation imbue top quarks with recoil momentum that propagate to charged leptons. This in turn leads to larger lepton momenta and thus larger invariant masses.

%%%%%%%%%%%%%%%%%%%%%%%%%%%%%%%%%%%%%%%%%%%%%%%%%%%%%%%%%%%%%%%%%%%%%%%%%%%%%%
\section{Outlook}\label{sec:outlook}
%%%%%%%%%%%%%%%%%%%%%%%%%%%%%%%%%%%%%%%%%%%%%%%%%%%%%%%%%%%%%%%%%%%%%%%%%%%%%%

Due to the large impact of QCD radiation at $\mathcal{O}(\alpha_s^4 \alpha)$ on the $\ttW$ process we are compelled to consider implications for other scenarios.  This includes, for example, inclusive $\ttZ$ production at the LHC, a situation that we now explore.

%%%%%%%%%%%%%%%%%%%%%%%%%%%%%%%%%%%%%%%%%%%%
\subsection{Inclusive \ttZ production at the LHC}\label{sec:outlook_ttZ}

\begin{figure}[!t]
\begin{center}
\includegraphics[width=\columnwidth]{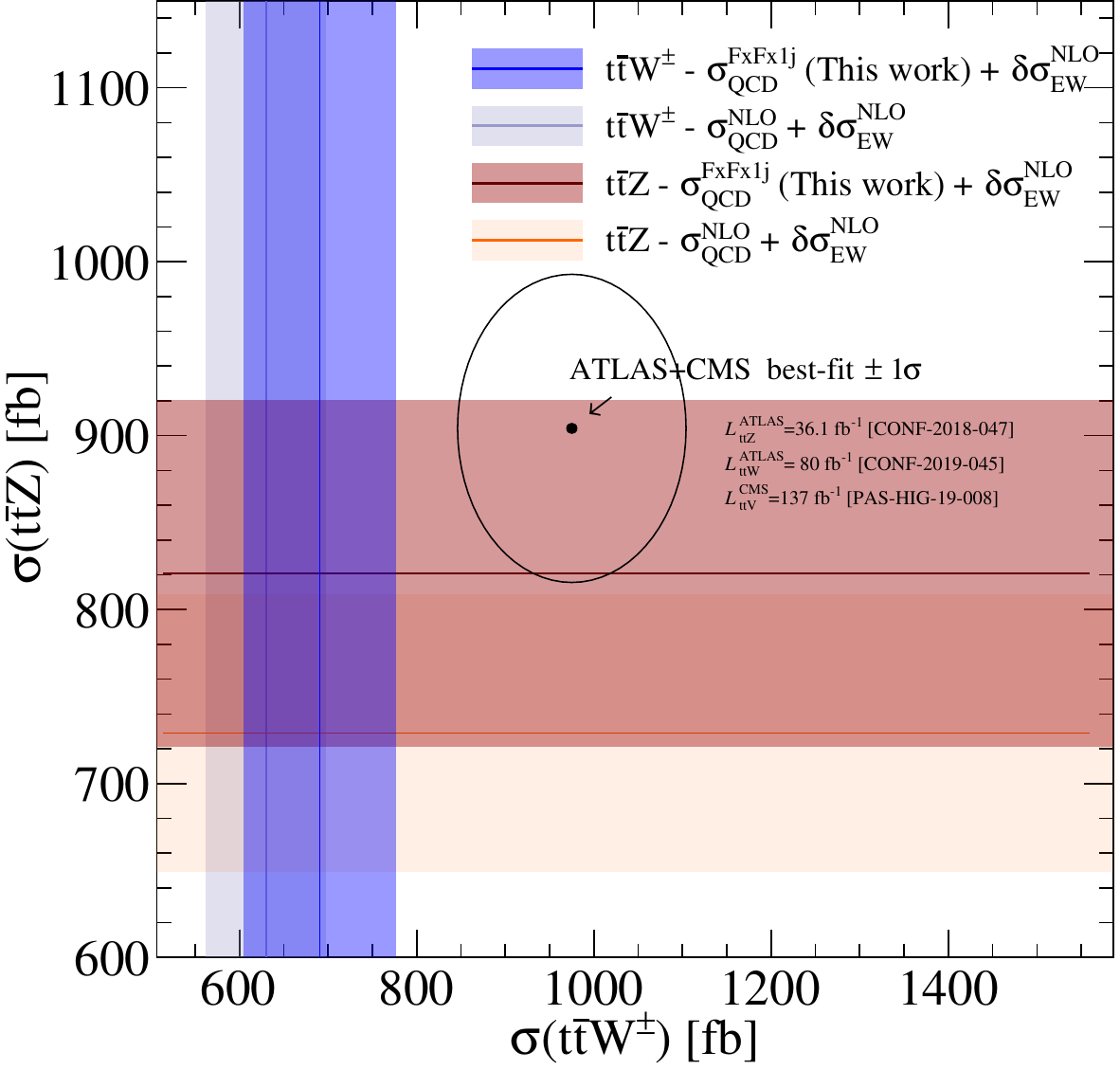}
\end{center}
\caption{
Central cross section predictions at $\sqrt{s}=13\TeV$ with $1\sigma$ uncertainty bands for the $\ttW$ and $\ttZ$ processes at NLO in QCD+EW (light) as well with FxFx matching (dark). Also shown are best-fit measurements from the ATLAS~\cite{ATLAS:2018ekx,ATLAS:2019nvo} and CMS~\cite{CMS:2020iwy} experiments.
}
\label{fig:ttWAnatomy_XSecOverview}
\end{figure}

The SM's SU$(2)_L$ gauge symmetry dictates that $\ttW$ and $\ttZ$ production are intimately related. However, due to differences in gauge quantum numbers, inclusive $\ttZ$ production occurs via different partonic channels, especially at lowest order. Hence, $\ttZ$ possesses a different sensitivity to QCD corrections.

To investigate these differences we repeat the work of section \ref{sec:inclusive} and report in the lower panel of table \ref{tab:normXSecFxFx} the cross sections for the $\ttZ$ rate at $\mathcal{O}(\alpha_s^2 \alpha)$ (LO), up to $\mathcal{O}(\alpha_s^3 \alpha)$ (NLO), and FxFx matching up to the first jet multiplicity (FxFx1j) for various inputs. Also reported are the scale and PDF uncertainties, as well as the QCD $K$-factor. We observe that QCD corrections generally impact the $\ttZ$ process in a comparable manner to the $\ttW$ process. More specifically, the NLO and nominal $(\dagger)$ FxFx1j $\ttZ$ rates possess the $K$-factors {$K_{\rm QCD}^{\ttZ}=1.48$ and $1.12$, respectively. In comparison, for $\ttW$ one finds respectively $K_{\rm QCD}^{\ttW}=1.57$ and $1.10$.} For other FxFx inputs we find comparable differences as in $\ttW$ but note that the $\ttZ$ FxFx1j rate suffers a slightly milder dependence on $p_T^{j \min}$ than \ttW.

To summarize our findings, we plot in figure \ref{fig:ttWAnatomy_XSecOverview} the cross section predictions with  uncertainties  at $\sqrt{s}=13\TeV$ for the $\ttW$ and $\ttZ$ processes at NLO in QCD+EW (light) and FxFx+EW (dark). (For \ttZ, we use the EW $K$-factor $K_{\rm NLO-EW} = 0.98$ \cite{Frixione:2015zaa}.) For theoretical uncertainties, we combine scale and PDF uncertainties in quadrature. Also shown is the error-weighted combination of  best-fit results from  ATLAS \cite{ATLAS:2018ekx,ATLAS:2019nvo} and CMS \cite{CMS:2020iwy}. Unlike the $\ttW$ case, we see  appreciable improvement in the agreement between the predicted and measured $\ttZ$ rate. However, like the $\ttW$ case, there remains a sizable theory uncertainty that prevents a more significant statement.

%%%%%%%%%%%%%%%%%%%%%%%%%%%%%%%%%%%%%%%%%%%%%%%%%%%%%%%%%%%%%%%%%%%%%%%%%%%%%%
\section{Summary and Conclusion}\label{sec:conclusions}
%%%%%%%%%%%%%%%%%%%%%%%%%%%%%%%%%%%%%%%%%%%%%%%%%%%%%%%%%%%%%%%%%%%%%%%%%%%%%%

In light of sustained measurements of an enhanced $t\overline{t}W$ cross section at the LHC, we report a systematic investigation into the role of the $\ttWj$ and $\ttWjj$ sub-processes in inclusive $\ttW$ production. We focus particularly on their impact on total and differential rate normalizations.

To conduct this study we revisited (see section \ref{sec:inclusive_state}) the state-of-the-art modeling for inclusive $\ttW$ production and took special note of estimated NNLO in QCD corrections that are employed in LHC analyses. Using LO and NLO in QCD computations, we then examined  (see section \ref{sec:inclusive_ttwjx})  the $\ttWj$ and $\ttWjj$ sub-processes. We report that a subset of real and virtual contributions at $\mathcal{O}(\alpha_s^4\alpha)$ in well-defined regions of phase space are positive and can arguably increase the inclusive $\ttW$ rate at NLO by at least {$10\%-14\%$}. Resummed results at NNLL do not suggest significant cancelations against pure two-loop diagrams.

Interestingly, using instead the non-unitary, NLO multi-jet matching scheme FxFx, we find (see section \ref{sec:inclusive_fxfx}) that these same QCD corrections at $\mathcal{O}(\alpha_s^4\alpha)$  increase the  inclusive $\ttW$ cross section at NLO by at most {$10\%-12\%$}. We attribute this difference to the treatment of soft logarithms in the matching scheme's Sudakov reweighting procedure. We obtain a slightly smaller central normalization for the inclusive $\ttW$ production rate than used in LHC analyses, which in turn  worsens slightly reported discrepancies. At the same time, after selection cuts and signal categorization, the FxFx description of $\ttW$ exhibits enhanced jet multiplicities that are slightly favored by data  (see section \ref{sec:differential}). 
Our main results are summarized in figure \ref{fig:ttWAnatomy_XSecOverview}. There we compare FxFx-improved cross sections for the $\ttW$ and $\ttZ$ processes to measurements at $\sqrt{s}=13\TeV$.

In conclusion, state-of-the-art
calculations do not obviously resolve existing tensions between SM predictions and LHC measurements of the $pp\to\ttW+X$ process. And while total and differential rates at the NLO multi-jet matching level provide an excellent description for many phenomena, for inclusive $\ttW$ and $\ttZ$ production the magnitude residual scale uncertainties, the moderate dependence on matching inputs, and the discrepancy with numerical estimates of $\mathcal{O}(\alpha_s^2)$ corrections motivate the need for a full description at NNLO in QCD.

%%%%%%%%%%%%%%%%%%%%%%%%%%%%%%%%%%%%%%%%%%%%%%%%%%%%%%%%%%%%%%%%%%%%%%%%%%%%%%
\section*{Acknowledgements}
We thank Carlos Fibonacci, Fabio Maltoni, Olivier Mattelaer, Aaron Vincent, and Marco Zaro for helpful discussions. We also thank Rikkert Frederix and Ioannis Tsinikos for their careful read of an earlier version of this manuscript.

This work was supported by the Faculty of Science Doctoral Internship Program at the University of the Witwatersrand. This work  has also received funding from the European Union's Horizon 2020 research and innovation programme as part of the Marie Skłodowska-Curie Innovative Training Network MCnetITN3 (grant agreement no. 722104), FNRS “Excellence of Science” EOS be.h Project No. 30820817.
SvB and BM are grateful for the continued support from the SA-CERN program that is supported by the National Research Foundation and the Department of Science and Innovation, and the Research Office of the University of the Witwatersrand.
RR is supported under the UCLouvain fund “MOVE-IN Louvain” and  acknowledge the contribution of the VBSCan COST Action CA16108.

Computational resources have been provided by the supercomputing facilities of the Universit\'e catholique de Louvain (CISM/UCL)  and the Consortium des \'Equipements de Calcul Intensif en F\'ed\'eration Wallonie Bruxelles (C\'ECI) funded by the Fond de la Recherche Scientifique de Belgique 
(F.R.S.-FNRS) under convention 2.5020.11 and by the Walloon Region.

%%%%%%%%%%%%%%%%%%%%%%%%%%%%%%%%%%%%%%%%%%%%%%%%%%%%%%%%%%%%%%%%%%%%%%%%%%%%%%
\bibliography{ttWAnatomy_refs}
%%%%%%%%%%%%%%%%%%%%%%%%%%%%%%%%%%%%%%%%%%%%%%%%%%%%%%%%%%%%%%%%%%%%%%%%%%%%%%%%%
\end{document}